\documentstyle{article}

\begin{document}

\baselineskip = 18pt

\thispagestyle{empty}
\rightline{LNF-97/003 (P)}
\rightline{gr-qc/9701065 \vspace{1cm}}
\begin{center}
{\bf Consistency of Orthodox Gravity}
 \vspace{1.5cm} \\
S. Bellucci\footnote{e-mail: bellucci@lnf.infn.it} \\
INFN-Laboratori Nazionali di Frascati, P.O. Box 13,
00044 Frascati, Italy
\vspace{1cm}\\
and
\vspace{1cm}\\
A. Shiekh\footnote{e-mail: shiekh@ictp.trieste.it} \\
International Centre for Theoretical Physics,
Strada Costiera 11, Trieste, Italy\vspace{2.5cm} \\
{\bf Abstract} 
\end{center}
A recent proposal for quantizing gravity is investigated for self
consistency. There are well-known difficulties in dealing with
Einstein gravity when resorting to the perturbative techniques of quantum
field theory. This however does not preclude the existence of a quantum
form. This Letter is all about such a subtle but important difference.
\vfill
\begin{center}
January 1997
\end{center}
\newpage

\section{Introduction}

The unification of gravity and the standard model describing strong
and electroweak interactions, thirty years after the formulation
of the latter, has yet to be achieved and is a formidable task.
Conceptual progress was made by introducing
an effective theory for processes at low energy (typically
less than the Planck mass \cite{wein1}). However, a fundamental theory of
gravity is still lacking. Our ignorance is parameterized by the
renormalized counter terms of the corresponding effective Lagrangian.
At very low energy, the theory is provided by general relativity. The natural
question to address is then the following: what about quantum corrections?
Donoghue tackled this issue \cite{d1} and we repeated his calculation
of the first quantum correction to the Newtonian potential \cite{newt}.
The validity of the abovementioned low-energy results holds,
irrespectively of any definite
proposal for the complete quantum theory of gravity.

In \cite{a1,a2} an approach to quantizing gravity
with a modified renormalization scheme was proposed, taking
a massive scalar field interacting with gravity as an example.
In this Letter we take up the issue of self consistency of a
candidate theory for the perturbative quantization of gravity.
For the sake of the simplicity, we focus on the
theory of self-interacting massive scalar fields coupled to quantum
gravity.

We plan this Letter in the following way.

We begin in Section 2 by describing the essential features of the
perturbative quantization of gravity. We start out with a maximal
gravity and then proceed to pin down physical criteria that
lead us to Einstein gravity quantized. We then discuss in
Section 3 the self consistency of this renormalized theory of quantum
gravity for a scalar field. We finish this Letter in
Section 4 by offering our remarks on the non-perturbative outlook.
\section{Orthodox Gravity: the perturbative approach}

A recent proposal for the perturbative quantization of gravity \cite{a1,a2}
suggests the quantization of an extended Lagrangian for
gravity, such that the counter terms are all accommodated
within the Lagrangian and so renormalization is formally 
achieved. In this Section, following and extending on this proposal,
we introduce Einstein gravity
quantized. Why the strange wording? Having granted that
Einstein gravity cannot be dealt with by resorting to the traditional
perturbative quantum field theoretical techniques, this does not
preclude the existence of a quantum form, and this Section deals
with this subtle but important difference.

It is well known that the abovementioned difficulties with the
quantization of Einstein gravity arise
for the simple reason that the infinities cannot be
accommodated within the starting Lagrangian. For the
purpose of illustration we will be discussing minimal
coupled gravity with massive scalar particles, as
governed by the Lagrangian:

\begin{equation}
{\cal L} = \sqrt {-g}
\left( {-2 \Lambda + R + {\textstyle{1 \over 2}}
g^{\mu \nu} ( \partial_\mu \phi )( \partial_\nu \phi ) +
{\textstyle{1 \over 2}} m^2 \phi^2}
\right)
\label{eq:starting}
\end{equation}
\\
The counter terms that carry the infinities cannot be
accommodated back within this starting Lagrangian, and
so the theory retains its divergent nature.

One often speaks of the starting Lagrangian as the
classical Lagrangian arguing that this is the starting
point for quantization, while the final Lagrangian,
which is of the same form, is referred to as the
quantized Lagrangian. This is a misleading notation,
as the original Lagrangian is divergent, having taken
up the counter terms, and the classical limit actually
arises in the $\hbar \to 0$ limit from the final complete
Lagrangian. This distinction will become especially poignant
in what follows.

Having noted that the difficulty in the quantization of minimal gravity
stemmed from the fact that the counter terms did not fall
back into the starting Lagrangian, one can resort to
extending the Lagrangian so as to ensure that the theory
is `formally' renormalizable. In this way we arrive at
what we can call maximal gravity.
This extended starting Lagrangian is constrained by
symmetry to be:

\begin{equation} L_0 = \sqrt{-g_0} 
\left(\matrix { -2\Lambda_0 + R_0 + 
\textstyle{1 \over 2} p_0^2 + 
\textstyle{1 \over 2} m_0^2 \phi_0^2 +
\textstyle{1 \over 4!} \phi_0^4
\lambda_0(\phi_0^2) + p_0^2 \phi_0^2
\kappa_0(\phi_0^2) + R_0 \phi_0^2 \gamma_0(\phi_0^2) \cr
\cr + p_0^4 a_0 (p_0^2,\phi_0^2) + R_0 p_0^2
b_0(p_0^2,\phi_0^2) + R_0^2 c_0(p_0^2,\phi_0^2) + R_{0\mu\nu}
R_0^{\mu\nu} d_0(p_0^2,\phi_0^2) + ... }
\right)
\end{equation}

\rightline{ \small \it 
(using units where $16 \pi G = 1$, $c = 1$)}

\noindent where $p_0^2$ is shorthand for $g_0^{\mu \nu}
\partial_\mu \phi_0 \partial_\nu \phi_0$ and not the
independent variable of Hamiltonian mechanics.
$\lambda_0$, $\kappa_0$, $\gamma_0$, $a_0$,
$b_0$, $c_0$, $d_0$ ... are arbitrary analytic functions, and
the second line carries all the higher derivative terms.
Strictly this is formal in having neglected gauge fixing and
the resulting presence of ghost particles.

Quantum anomalies
arise from a conflict between symmetries, where only one can be
maintained \cite{Mann}. For this reason no such trouble is
anticipated here.
At this point some remarks are in order, to clarify
why there is no trouble with a gravitational anomaly
in our case. We are a bit cavalier on this point, but anomalies do not
just turn up to break a symmetry (one can always fix things, so that the
symmetry is restored).
The trouble arises when there are two symmetries (say the conformal
one also). Then one can restore any symmetry, {\it except} the fixing
is different for each symmetry, and the two fixings tend to be
in conflict. One could, for example, get rid of the conformal anomaly
in massless gravity, but at the price of a worse anomaly.
All this is detailed in Mann's review \cite{Mann}.

The price for having achieved `formal' renormalization, is
that the theory (with its infinite number of arbitrary
renormalized parameters) has only a limited
predictive content.
As explained in \cite{d1}, the measurement of a finite number of parameters
is needed, in order to make a class of predictions at a given
accuracy for a specified energy scale.
The failure to quantize has been rephrased from a problem of
non-renormalizability to a difficulty in the predictability.

Despite this, after renormalization we are led to:

\begin{equation} L = \sqrt{-g} 
\left(\matrix { -2\Lambda + R + 
\textstyle{1 \over 2} p^2 + 
\textstyle{1 \over 2} m^2 \phi^2 +
\textstyle{1 \over 4!} \phi^4 \lambda(\phi^2) + p^2 \phi^2
\kappa(\phi^2) + R \phi^2 \gamma(\phi^2) \cr \cr + p^4 a
(p^2,\phi^2) + R p^2 b(p^2,\phi^2) + R^2 c(p^2,\phi^2) +
R_{\mu\nu} R^{\mu\nu} d(p^2,\phi^2) + ... }
\right)
\end{equation}
\\
However, there remain physical criteria to pin down
some of these arbitrary factors. Since in general (except for
special cases) the higher
derivative terms lead to acausal classical behavior, their
renormalized coefficient can be put down to zero on physical
grounds. This is however not always correct, for there exist particular
cases of Lagrangians having higher derivative terms, where the ghost
poles cancel, the simplest such example being the Gauss-Bonnet term.
Hence, there are higher derivative lagrangians which do not introduce
propagating Weyl ghosts and so do not spoil unitarity.
The corresponding terms in the action yield a topological invariant,
whose metric variation vanishes identically, so one is left (in 4
dimensions) with a smaller number of arbitrary renormalized
coefficients,
whose value can only be determined from experiments. To quote
from the book by Birrell-Davies \cite{a10} (page 162): ``In principle there
is no reason why these renormalized quantities may not be set equal to
zero, thus recovering Einstein's theory. Quantum field theory merely
indicates that terms involving higher derivatives of the metric
are a priori expected.'' The simplest approach is to force the
corresponding renormalized coefficients to vanish.

This still leaves the three arbitrary functions:
$\lambda(\phi^2)$, $\kappa(\phi^2)$ and
$\gamma(\phi^2)$, associated with the terms $\phi^4$,
$p^2 \phi^2$, and
$R \phi^2$ respectively. The last may be abandoned on the
grounds that its coefficient must be extremely small,
not to defy the equivalence principle. To see this,
begin by considering the first term of the Taylor expansion,
namely $R\phi^2$;
this has the form of a mass term, hence one would be able to make local
measurements of mass to determine
the curvature, and so contradict the equivalence principle
(charged particles, with their non-local fields, have this term
present with a fixed coefficient).\footnote{Tests of the equivalence
principle and bounds on the deviations
from Newton's inverse square law constrain also the antigravity fields
advocated in $N=2,8$ supergravity \cite{bf1,bf2}. For a discussion of
possible violations of the equivalence principle, see also \cite{a3}.}
One could of course retain such terms, at the price of abandoning
the equivalence principle, with a coefficient small enough,
to be tolerable from the observational standpoint. We take, as
explained above, the simplest approach. The same line of reasoning
applies to the remaining terms, $R\phi^4$, $R\phi^6$, ... etc.

This leaves us the two remaining infinite families of
ambiguities with the terms $\phi^4\lambda(\phi^2)$ and
$p^2\phi^2\kappa(\phi^2)$. In the limit of flat space in 3+1
dimensions this will reduce to a renormalized theory in the
traditional sense if $\lambda(\phi^2)=constant$, and
$\kappa(\phi^2)=0$. So one is led to proposing that the
physical parameters should be:\footnote{Our approach does not make
necessary to set the cosmological constant to zero. One can perhaps advocate
only empirical reasons for doing so. An explanation of the
shadowing of the cosmological constant effect by those of newtonian
gravity has been given recently
for superconformal invariant theories \cite{b}.}
\begin{equation}
\matrix {
\kappa(\phi^2) = \gamma(\phi^2) = 0 \cr \cr
a(p^2,\phi^2) = b(p^2,\phi^2) = c(p^2,\phi^2) = d(p^2,\phi^2)
=... =0 \cr \cr
\lambda(\phi^2) = \lambda = {\it scalar\ particle\ self\
coupling\ constant} \cr \cr m = {\it mass\ of\ the\ scalar\
particle} }
\end{equation}
\\
\noindent and so the renormalized theory of quantum gravity
for a scalar field will have the Einstein form:

\begin{equation} L = \sqrt{-g} 
\left( -2\Lambda + R + 
\textstyle{1 \over 2} p^2 + 
\textstyle{1 \over 2} m^2 \phi^2 +
\textstyle{1 \over 4!} \lambda \phi^4
\right)
\label{eq:final}
\end{equation}
\\
This is a candidate for the long sought after Einstein gravity
quantized, and not quantized Einstein gravity; and the classical
theory arises in the $\hbar \to 0$ limit of this.

\section{Self-consistent gravity}

One might now worry about the renormalization group pulling
the coupling constants around.

Since we are interested only that the zeroed couplings
remain so, we shall name them as external couplings,
in so much as they belong to terms outside the final
renormalised Lagrangian (eq.~\ref{eq:final}).
The finite number remaining  will naturally take up the
designation of internal couplings.

When
a coupling runs, its value at some energy scale must be specified.
The beta functions then determines how the coupling varies for
other scales. It can now be seen to be a trivial matter to stop
the external couplings from running, namely by zeroing them at
an infinite scale.

Notice that in the previous Section the coupling
$G$ was not omitted, but set to unity. This is
a matter of convenience and is not supposed to suggest that
$G$ cannot run. So, strictly $G$ must be restored before
embarking in any future calculation, a thing we are presently
doing (with $G$ in the Feynman rules).

\section{Non-perturbative perspective}

The above argument was done completely within a perturbative
context, and one might wonder if a non-perturbative perspective
would lead to the same proposal, and then perhaps without the
infinities of the perturbative approach.

\subsection{Ashtekar variables approach}

We wish here to just point out that there exists a very dynamic
program, i.e. the Ashtekar variables approach \cite{av}, trying precisely to
quantize gravity in a non-perturbative fashion.\footnote{This approach
is not limited to the non-perturbative investigation.} It is at this point
yet unclear, whether Einstein gravity can be quantized along these lines.

\subsection{The high tension string}

String theory might be thought of as another attempt
to quantize gravity by generalizing away from point
particle theory (supergravity having failed).

When viewing string theory as a higher derivative,
infinitely large Lagrangian, one sees many similarities
with orthodox gravity, excepting that string theory
has only one, and not an infinity, of extra parameters
in the form of the string tension.
It then becomes very natural to wonder about the
point particle limit of the superstring, when one
anticipates the appearance of supergravity. There
immediately arises a question of how to resolve
the fact that supergravity is not renormalizable
(at least using traditional quantum field theory techniques), but
that the string in the high tension limit exists.
The above investigation makes the resolution rather
transparent in so much as the starting Lagrangian
is not that of supergravity even in the limit,
for one has no reason to suppose the higher derivative
bare terms disappear in the high tension limit.
Again, only the theory quantized reduces to supergravity.

In this way one might view orthodox gravity as the
point particle (high tension) limit of the string,
and as such it is a second confirmation of the existence
of orthodox gravity as a candidate for gravity quantized.

\subsection{Occam's gravity}

We generalised to supergravity and string theory because
our former candidates failed to give us quantum gravity.
But why go to the complexities of higher dimensions, or
a new set of particles if we can locate a simpler candidate?
Naturally, at the end of the day, it is not a choice for
us to make, but rather a question to be put to nature.

We are aware of the fact that traditional power counting
renormalizability can be replaced \cite{w} by a less restrictive
one, where gravitational metric theories are considered as
renormalizable. The appearance of the Einstein-Hilbert action
within this framework is due to the suppression of higher derivative
terms by powers of (possibly) the Planck mass.
We paid special attention to the issue of unitarity,
which is often lost for higher derivative gravity. We
stress that we are not truncating the higher derivative gravity
Lagrangian, hence unitarity is not lost in our approach. In this sense
we have gone beyond the approximation of the effective field
theory approach, where higher order corrections are suppressed.

It is rather paradoxical that we have arrived at a minimalist
proposal by having first resorted to a maximal theory.
But it is satisfying in having added every ingredient to
the broth and seeing it slim itself down on its
own accord.

Being such a simple candidate, one can immediately go about
calculating with this proposal.

\section*{Acknowledgement}

Thanks are due to the referees of our Letter for useful comments.
This work does not reflect the views of the High Energy
Physics group at ICTP.


\end{document}